\documentstyle[eqsecnum,aps,epsf,
twocolumn,feynmf]{revtex}
\newcommand{\mean}[1]{\langle #1 \rangle_{m^2}}
\newcommand{\cum}[1]{\langle #1 \rangle_{m^2,c}}
\def\setval{\fmfset{wiggly_len}{2mm}\fmfset{arrow_len}{2mm}
\fmfset{dash_len}{1.5mm}\fmfpen{0.125mm}\fmfset{dot_size}{1thick}}
\begin{document}
\title{Strong-Coupling Calculation of  Fluctuation
Pressure of a \\Membrane Between Walls}
\author{M. Bachmann\thanks{Supported
by the Studienstiftung des deutschen Volkes.}
\thanks{email: mbach@physik.fu-berlin.de},
H. Kleinert\thanks{email: kleinert@physik.fu-berlin.de}, and
A. Pelster\thanks{email: pelster@physik.fu-berlin.de}}
\address{Institut f\"ur Theoretische Physik, Freie Universit\"at Berlin,
Arnimallee 14, 14195 Berlin}
\date{\today}
\maketitle
\begin{abstract}
We calculate analytically the proportionality constant in the pressure law
of a membrane
between parallel walls from the strong-coupling limit of
variational perturbation theory up to third
order. Extrapolating the zeroth to third
 approximations
to infinity yields the pressure constant
$\alpha=0.0797149$. This result lies well within
the error bounds of the most accurate available Monte-Carlo result
$\alpha^{\rm MC}=0.0798\pm 0.0003$.
\end{abstract}
\section{Membrane Between Walls}
The violent thermal out-of-plane fluctuations of a membrane
between parallel walls generate a pressure $p$ following the law
\begin{equation}
  \label{mem01}
  p=\alpha\frac{k_B^2 T^2}{\kappa (d/2)^3},
\end{equation}
whose form was first derived by Helfrich~\cite{helfrich1}.
Here, $\kappa$ denotes the elasticity constant of the membrane,
and $d$ the distance between the walls.
The exact value of the prefactor $\alpha$ is unknown, but
estimates have been derived from crude theoretical approximations
by Helfrich~\cite{helfrich1} and by Janke and Kleinert~\cite{jk} which
yielded
\begin{equation}
\label{mem01b}
\alpha_{\rm H}^{\rm th}\approx 0.0242,\quad
\alpha_{\rm JK}^{\rm th}\approx 0.0625.
\end{equation}
More precise values were found from Monte-Carlo simulations by
Janke and Kleinert~\cite{jk} and by Gompper and Kroll~\cite{mc} which gave
\begin{equation}
  \label{mem01c}
  \alpha_{\rm JK}^{\rm MC}\approx 0.079\pm 0.002,\quad
\alpha_{\rm GK}^{\rm MC}\approx 0.0798\pm 0.0003.
\end{equation}
In a previous work~\cite{kl277}, a systematic method was developed for
calculating $\alpha$ with any desired high accuracy. Basis
for this method is the strong-coupling version of variational perturbation
theory~\cite{kleinertPI}. In that theory, the free energy of the membrane is
expanded into a sum of connected loop dia\-grams, which is eventually taken
to infinite coupling strength to account for the hard walls. As a first
approximation, an infinite set of dia\-grams was calculated, others were
estimated by invoking a mathematical analogy with a similar one-dimensional
system of a quantum mechanical particle between walls.
The result of this procedure was a pressure constant
\begin{equation}
  \label{mem01cc}
  \alpha_{K}^{\rm th}=\frac{\pi^2}{128}= 0.0771063\ldots,
\end{equation}
very close to (\ref{mem01c}).

It is the purpose of
this paper to go beyond this estimate by calculating all dia\-grams
up to four loops exactly. In this way, we improve the analytic
approximation (\ref{mem01cc}) and obtain a value
\begin{equation}
  \label{mem01ccc}
  \alpha^{\rm th}\approx 0.0797149,
\end{equation}
which is in excellent agreement with the precise
MC value $\alpha_{\rm GK}^{\rm MC}$ in Eq.~(\ref{mem01c}).
\section{The Smooth Potential Model of the Membrane Between Walls}
To set up the theory, we let the membrane lie in the ${\bf x}$-plane
and fluctuate in the $z$-direction with vertical displacements
$\varphi({\bf x})$. The walls at $z=\pm d/2$ restrict
the displacements to the interval $\varphi\in (-d/2,d/2)$. Near zero
temperature, the
thermal fluctuations are small, $\varphi({\bf x})\approx 0$.
 The curvature energy $E$ of the membrane has the harmonic
approximation~\cite{helfrich1}
\begin{equation}
  \label{mem01d}
  E=\frac{1}{2}\kappa\int dx^2\,[\partial^2 \varphi({\bf x})]^2.
\end{equation}
The thermodynamic partition function $Z$ of the membrane is
given by the sum over all Boltzmann factors of field
configurations $\varphi({\bf x})$
\begin{eqnarray}
  \label{mem02}
  Z=\prod\limits_{\bf x}&&\left[\int_{-d/2}^{+d/2} \frac{d\varphi({\bf x})}
{\sqrt{2\pi k_B T/\kappa}}\right]\nonumber\\
&&\times\exp\left\{-\frac{\kappa}{2k_B T}
\int d^2x\,[\partial^2 \varphi({\bf x})]^2 \right\}.
\end{eqnarray}
This simple harmonic functional integral poses the
problem of dealing with a finite range of fluctuations. This problem is
solved by the strong-coupling theory of Ref.~\cite{kl277}
as follows.

If the area of the membrane is denoted by $A$, the partition function
(\ref{mem06}) determines the free energy per area as
\begin{equation}
  \label{mem07b}
  f=-\frac{1}{A}\,{\rm ln}\,Z.
\end{equation}
By differentiating $f$ with respect to the distance $d$ of the walls,
we obtain the pressure $p=-\partial f/\partial d$.
\subsection{Smooth Potential Adapting Walls}
We introduce some smooth potential restricting the fluctuations
$\varphi({\bf x})$ to the interval $(-d/2,d/2)$, for instance
\begin{eqnarray}
  \label{mem03}
  V(\varphi({\bf x}))&=&m^4\frac{d^2}{\pi^2}\tan^2 \frac{\pi}{d}
\varphi({\bf x})\nonumber\\
&\equiv& m^4\varphi^2({\bf x})+\frac{\pi^2}{d^2}
V_{\rm int}(\varphi({\bf x})),
\end{eqnarray}
where we have split the potential into harmonic and interacting part
\begin{eqnarray}
  \label{mem04}
  V_{\rm int}(\varphi({\bf x}))=m^4\Big[& &\varepsilon_4\varphi^4({\bf x})+
\varepsilon_6\left(\frac{\pi}{d}\right)^2\varphi^6({\bf x})+\nonumber\\
&&\varepsilon_8\left( \frac{\pi}{d}\right)^4\varphi^8
({\bf x})+\ldots\Big]
\end{eqnarray}
with $\varepsilon_4=2/3, \varepsilon_6=17/45, \varepsilon_8=62/315,\ldots$ .
Thus we are left with the functional integral
\begin{eqnarray}
  \label{mem06}
  Z=\oint{\cal D}\varphi({\bf x})\,\exp\Big(&&-\frac{1}{2}\int
d^2x\,\Big\{[\partial^2 \varphi({\bf x})]^2 +m^4\varphi^2({\bf x})+\nonumber\\
&&\frac{\pi^2}{d^2}V_{\rm int}(\varphi({\bf x}))\Big\}\Big),
\end{eqnarray}
where we have set $\kappa=k_B T=1$. After truncating the Taylor expansion
around the origin,
the periodicity of the trigonometric function is lost and the integrals
over $\varphi({\bf x})$ in (\ref{mem02}) can be taken from
$-\infty$ to $+\infty$. The interacting part is treated perturbatively.
Then, the harmonic part of $V(\varphi(\bf x))$ leads to an exactly integrable
partition function $Z_{m^2}$.
The mass parameter $m$ is arbitrary at the moment, but will eventually
taken to zero, in which case the potential $V(\varphi(\bf x))$ describes
two hard walls at $\varphi=\pm d/2$.

We shall now calculate a perturbation expansion for $Z$ up to
four loops. This will serve as a basis for the limit $m\to 0$, which
will require the strong-coupling theory of Ref.~\cite{kl277}.
\subsection{Perturbation Expansion for Free Energy}
The perturbation expansion proceeds from the harmonic part of
Eq.~(\ref{mem05b}):
\begin{eqnarray}
\label{mem08}
Z_{m^2}=\oint{\cal D}\varphi({\bf x})\,e^{-{\cal A}_{m^2}[\varphi]/2}
=e^{-A f_{m^2}}.
\end{eqnarray}
with
\begin{equation}
  \label{mem08aa}
  {\cal A}_{m^2}[\varphi]=\int d^2x\,\left\{[\partial^2 \varphi({\bf x})]^2
 +m^4\varphi^2({\bf x})\right\}.
\end{equation}
{}From Refs.~\cite{jk,kl277}, the harmonic free energy per unit area $f_{m^2}$
is known as
\begin{equation}
  \label{mem08b}
  f_{m^2}=\frac{1}{8}m^2.
\end{equation}
The harmonic correlation functions associated with (\ref{mem08}) are
\begin{eqnarray}
  \label{mem09}
&&\mean{O_1(\varphi({\bf x}_1))\,O_2(\varphi({\bf x}_2))\,\cdots}=
\nonumber\\
&&\frac{1}{Z_{m^2}}
\oint{\cal D}\varphi({\bf x})O_1(\varphi({\bf x}_1))\,
O_2(\varphi({\bf x}_2))\cdots e^{-{\cal A}_{m^2}[\varphi]/2}
\end{eqnarray}
where the functions $O_i(\varphi({\bf x}_j))$ may be arbitrary polynomials of
$\varphi({\bf x}_j)$. The basic harmonic correlation function
\begin{equation}
  \label{mem10}
  G_{m^2}({\bf x}_1,{\bf x}_2)=\mean{\varphi({\bf x}_1)\,
\varphi({\bf x}_2)},
\end{equation}
determines, by Wick's rule, all correlation functions (\ref{mem09}) as sums
of products of (\ref{mem10}):
\begin{eqnarray}
  \label{mem11}
&&  \mean{\varphi({\bf x}_1)\cdots\varphi({\bf x}_n)}=\nonumber\\
&&\hspace{2mm}\sum\limits_{\rm pairs}\,
G_{m^2}({\bf x}_{P(1)},{\bf x}_{P(2)})\cdots
G_{m^2}({\bf x}_{P(n-1)},{\bf x}_{P(n)}),
\end{eqnarray}
where the sum runs over all pair contractions, and $P$ denotes the associated
index permutations. The harmonic correlation function  (\ref{mem10})
is in momentum space
\begin{equation}
  \label{mem12}
  G_{m^2}({\bf
k})=\frac{1}{k^4+m^4}=\frac{i}{2m^2}\left[\frac{1}{k^2+im^2}-\frac{1}{k^2-im^2}
\right],
\end{equation}
thus being proportional to the difference of two ordinary correlation
functions $(p^2-\mu^2)^{-1}$ with an imaginary square mass
$\mu^2=\pm i m^2$.
{}From their known ${\bf x}$-space form
we have immediately
\begin{eqnarray}
  \label{mem13}
&&  G_{m^2}({\bf x}_1,{\bf x}_2)=G_{m^2}({\bf x}_1-{\bf x}_2)\nonumber\\
&&=\frac{i}{4\pi m^2}
\left[K_0(\sqrt{i}m|{\bf x}_1-{\bf x}_2|)-
K_0(\sqrt{-i}m|{\bf x}_1-{\bf x}_2|)\right],\nonumber\\
\quad
\end{eqnarray}
where $K_0(z)$ is a modified Bessel function~\cite[Sec.~8.432]{gradstein1}.
At zero distance, the ordinary harmonic correlations are logarithmically
divergent,
but the difference is finite yielding $G_{m^2}(0)=1/8 m^2$.

We now expand the  partition function (\ref{mem06}) in powers
of $gV_{\rm int}(\phi({\bf x}))$, where $g\equiv\pi^2/d^2$, and
use (\ref{mem09}) to obtain a perturbation series for $Z$.
Going over to the cumulants, we find the free energy per unit area
\begin{eqnarray}
  \label{mem16}
&&f=f_{m^2}+\frac{g}{2A}\int d^2x\cum{V_{\rm int}(\varphi({\bf x}))}-
\nonumber\\
&&\frac{g^2}{2!}\frac{1}{4A}\int d^2x_1 d^2x_2
\cum{V_{\rm int}(\varphi({\bf x}_1))V_{\rm int}(\varphi({\bf x})_2)}+
\ldots,\nonumber\\
\quad
\end{eqnarray}
where the subscript $c$ indicates the cumulants. Inserting the expansion
(\ref{mem04}) and using (\ref{mem09}) and (\ref{mem11}), the series can be
written as
\begin{equation}
  \label{mem17}
  f=m^2\left[a_0+\sum\limits_{n=1}^\infty\,a_n\left(\frac{g}{m^2} \right)
^{n}\right],
\end{equation}
where the coefficients $a_n$ are dimensionless real numbers, starting with
$a_0=1/8$.
The higher expansion coefficients $a_n$ are combinations of integrals
over the connected correlation functions:
\begin{eqnarray}
  \label{mem18b}
a_1&=&\varepsilon_4\frac{m^2}{2A}\int d^2x\,
\cum{\varphi^4({\bf x})},\\
\label{mem18c}
a_2&=&\varepsilon_6\frac{m^6}{2A}\int d^2x\,
\cum{\varphi^6({\bf x})}\nonumber \\
&-&\varepsilon_4^2\frac{m^{10}}{8A}\int d^2x_1d^2x_2\,
\cum{\varphi^4({\bf x}_1)
\,\varphi^4({\bf x}_2)},\\
\label{mem18d}
a_3&=&\varepsilon_8\frac{m^8}{2A}\int d^2x\,\cum{v^8({\bf x})}
\nonumber \\
&-&\varepsilon_4\varepsilon_6\frac{m^{12}}{4A}\int d^2x_1d^2x_2\,
\cum{\varphi^6({\bf x}_1)
\,\varphi^4({\bf x}_2)}\\
&+&\varepsilon_4^3\frac{m^{16}}{48A}\int d^2x_1d^2x_2
d^2x_3\cum{\varphi^4({\bf x}_1)\varphi^4({\bf x}_2)
\varphi^4({\bf x}_3)},\nonumber
\end{eqnarray}
To find the free energy (\ref{mem17}) between walls, we must go to
the limit $m^2\to 0$.
Following~\cite{kleinertPI,kl277}, we substitute $m^2$ by the variational
parameter $M^2$, which is introduced via the trivial identity
$m^2\equiv\sqrt{M^4-g r}$ with $r = (M^4-m^4)/g$,
and expand this in powers of $g$ up to the order $g^N$.
In the limit $m^2\to 0$, this expansion reads
\begin{equation}
  \label{mem05b}
  m^2(M^2)=M^2-\frac{1}{2}\frac{r}{M^2}g-
\frac{1}{8}\frac{r^2}{M^6}g^2-\frac{1}{16}
\frac{r^3}{M^{10}}g^3-\ldots\:.
\end{equation}
Inserting this into (\ref{mem17}), reexpanding in powers of $g$,
and truncating after the $N$th term, we arrive at the
free energy per unit area
\begin{equation}
  \label{mem20}
  f_N(M^2,d)=M^2\,a_0\, b_0+\sum\limits_{n=1}^N\,a_n
g^nM^{2(1-n)}b_n
\end{equation}
with
\begin{equation}
  \label{mem21}
  b_n=\sum\limits_{k=0}^{N-n}\,(-1)^k\,\left(\begin{array}{c}(1-n)/2 \\ k
 \end{array}\right)
\end{equation}
being the binomial expansion of $(1-1)^{(1-n)/2}$ truncated after the
$(N-n)$th term~\cite{kl277}.
The optimization of (\ref{mem20}) is done as
usual~\cite{kleinertPI} by
determining the minimum of $f_N(M^2,d)$ with respect to the variational
parameter $M^2$, i.e. by the condition
\begin{equation}
  \label{mem22}
  \frac{\partial f_N(M^2,d)}{\partial M^2}\stackrel{!}{=}0,
\end{equation}
whose solution gives the optimal value $M^2_N(d)$. Resubstituting this
result into Eq.~(\ref{mem20}) produces the optimized free energy
$f_N(d)=f_N(M^2_N(d),d)$, which only depends on the distance
as $f_N(d)=4 \alpha_N/d^2$. Its
derivative with respect to $d$ yields the
desired pressure law with the $N$th order approximation for the constant
$\alpha_N$:
\begin{equation}
  \label{mem23}
  p_N=\alpha_N\left(\frac{d}{2}\right)^{-3}.
\end{equation}
We must now calculate the
cumulants occuring in the expansion (\ref{mem17}).
\section{Evaluation of the Fluctuation Pressure up to Four-Loop Order}
The correlation functions appearing in (\ref{mem18b})--(\ref{mem18d}) are
conveniently represented by Feynman graphs. Green functions are pictured as
solid lines and local
interactions as dots, whose
coordinates are integrated over:
\setlength{\unitlength}{1mm}
\begin{fmffile}{graph}
\begin{eqnarray}
\label{mem24}
\parbox{20mm}{\centerline{
\begin{fmfgraph*}(7,3)
\setval
\fmfleft{i}
\fmfright{o}
\fmf{plain}{i,o}
\fmflabel{${\bf x}_1$}{i}
\fmflabel{${\bf x}_2$}{o}
\end{fmfgraph*}
}}&\equiv& G_{m^2}({\bf x}_1,{\bf x}_2), \\
\label{mem25}
\parbox{20mm}{\centerline{
\begin{fmfgraph*}(3,3)
\setval
\fmfforce{0.5w,0.5h}{v}
\fmfdot{v}
\end{fmfgraph*}
}}&\equiv&\int d^2x.
\end{eqnarray}
These rules can be taken over to momentum space in the usual way. One easily
verifies that the integrals over the connected correlation functions in
(\ref{mem18b})--(\ref{mem18d}) have a dimension $A/m^{2(n+V-1)}$,
where $V$ is the number of the vertices of the associated Feynman dia\-grams.
Thus we parametrize each Feynman dia\-gram by $v A/m^{2(n+V-1)}$, with a
dimensionless number $v$, which includes the multiplicity.
In Tab.~\ref{tab1}, we have listed the values $v$ for all dia\-grams up to
four loops. No divergences are encountered. Exact results are stated as
fractional numbers.
The other numbers are obtained by
numerical integration, which are reliable up to the last written digit.
The right-hand column shows numbers $v_{\rm K}$ obtained by the earlier
approximation~\cite{kl277}, where all the Feynman dia\-grams were
estimated by an analogy to the the problem of a particle in a box.
In Ref.~\cite{kl277}, it was shown that the value $v$ of a large class of
dia\-grams of the
membrane problem can be obtained by simply dividing the value of the
corresponding particle-in-a-box-dia\-gram $v_{\rm PB}$ by a factor $1/4^L$,
where $L$ is the number of loops in the dia\-grams.

Inserting the numbers in Tab.~\ref{tab1} into
(\ref{mem18b})--(\ref{mem18d}), we obtain
the coefficients $a_1,a_2,a_3$
of the free energy per area (\ref{mem20}), which is then extremized in $M^2$.
To see how the results evolve
from order to order, we start with the first order
\begin{equation}
\label{mem26}
f_1(M^2,d)=\frac{1}{2}a_0M^2+a_1\frac{\pi^2}{d^2}
\end{equation}
with $a_0=1/8$ and $a_1=1/64$. Here, an optimal value of $M^2$ does not
exist. Thus we simply use the
perturbative result for $m=0$ which is equal to (\ref{mem26}) for $M=0$.
Differentiating $f_1(0,d)$ with respect to $d$ yields the pressure constant
in (\ref{mem23}):
\begin{equation}
  \label{mem27}
  \alpha_1=\frac{1}{4}a_1\frac{\pi^2}{d^2}=\frac{\pi^2}{256}\approx 0.038553.
\end{equation}
This value is about half as big as the Monte-Carlo estimates (\ref{mem01c})
and agrees with the value found in~\cite{kl277}.
To second order, the reexpansion (\ref{mem20}) reads
\begin{equation}
  \label{mem28}
  f_2(M^2,d)=\frac{3}{8}a_0M^2+a_1\frac{\pi^2}{d^2}+
a_2\frac{\pi^4}{d^4}\frac{1}{M^2}
\end{equation}
with $a_2\approx 1.0882\cdot 10^{-3}$ from Tab.~\ref{tab1}.
Minimizing this energy in $M^2$ yields an optimal value
\begin{equation}
\label{mem28b}
M^2_2(d)=\sqrt{\frac{8}{3}\frac{a_2}{a_0}}\frac{\pi^2}{d^2}\approx 0.152362
\frac{\pi^2}{d^2},
\end{equation}
and
\begin{equation}
  \label{mem28c}
  f_2(d)=\frac{\pi^2}{d^2}\left(a_1+\sqrt{\frac{3}{2}a_0a_2} \right).
\end{equation}
Inserting $a_0=1/8$ and $a_1,a_2$ from Tab.~\ref{tab1}, we obtain
\begin{equation}
  \label{mem29}
  \alpha_2\approx 0.073797,
\end{equation}
thus improving drastically the first-order estimate (\ref{mem27})
This value is by a factor $1.026$ larger than that obtained in the
approximation of Ref.~\cite{kl277}.

Continuing this proceeding to third order, we must minimize
\begin{equation}
  \label{mem30}
  f_3(M^2, d)=\frac{5}{16}a_0M^2+a_1\frac{\pi^2}{d^2}+\frac{3}{2}a_2
\frac{\pi^4}{d^4}\frac{1}{M^2}+a_3\frac{\pi^6}{d^6}\frac{1}{M^4}
\end{equation}
with $a_3\approx 2.7631\cdot 10^{-5}$.
The optimal value of $M^2$ is
\begin{eqnarray}
\label{mem30b}
M^2_3(d)&=&\sqrt{\frac{32}{5}\frac{a_2}{a_0}}\,\cos\left[\frac{1}{3}\arccos
\sqrt{\frac{5}{2}\frac{a_0a_3^2}{a_2^3}}\right]\frac{\pi^2}{d^2}\nonumber\\
&\approx& 0.219608\frac{\pi^2}{d^2}.
\end{eqnarray}
Inserted into (\ref{mem30}), we find the four-loop
approximation for the proportionality constant $\alpha$:
\begin{equation}
  \label{mem31}
  \alpha_3\approx 0.079472.
\end{equation}
This result is in very good agreement of the Monte-Carlo
results in (\ref{mem01c}). It differs from the approximate value of the method
presented in Ref.~\cite{kl277} by a factor $1.047$.

An even better result will now be obtained by extrapolating the sequence
$\alpha_1,\alpha_2,\alpha_3$ to infinite order.
\section{Extrapolation Towards the Exact Constant}
Variational perturbation theory exhibits typically an exponentially
fast convergence.
This was exactly proven for the anharmonic oscillator~\cite{kleinertPI}.
Other systems treated by variational perturbation theory
show a similar behavior~\cite{kuerz}.
Assuming that an exponential convergence exists
also here, we may extrapolate the sequence of values
$\alpha_1,\alpha_2,\alpha_3$ calculated above to infinite order. It is
useful to extend this sequence by one more value at the lower end,
$\alpha_0=0$, which follows from the one-loop energy (\ref{mem08b}) at
$m^2=0$.
This sequence is now extrapolated towards a hypothetical exact value
$\alpha_{\rm ex}$ by parametrizing the approach as
\begin{equation}
  \label{mem33}
  \alpha_{\rm ex}-\alpha_N=\exp\left(-\eta-\xi N^\epsilon\right),
\end{equation}
The parameters
$\eta$, $\xi$, $\varepsilon$, and the unknown value of $\alpha_{\rm ex}$ are
determined from the four values $\alpha_0,\ldots,\alpha_3$, with the result
\begin{equation}
  \label{mem34}
  \eta=2.529298,\quad \xi=0.660946,\quad \epsilon=1.976207,
\end{equation}
and the extrapolated value for the exact constant:
\begin{equation}
  \label{mem35}
  \alpha_{\rm ex}=0.0797149.
\end{equation}
This is now in perfect agreement with the Monte-Carlo values (\ref{mem01c}).

\begin{figure}
\centerline{
\setlength{\unitlength}{1cm}
\begin{picture}(8.5,6)
\put(0,0){\makebox(8.5,6){\epsfxsize=8cm \epsfbox{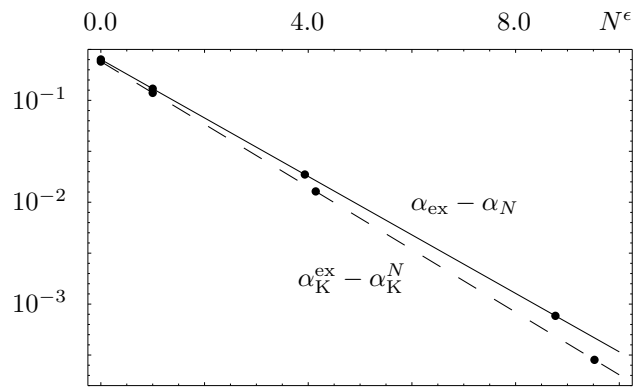}}}
\put(0.95,5.6){$0.0$}
\put(3.7,5.6){$4.0$}
\put(6.45,5.6){$8.0$}
\put(7.8,5.6){$N^\epsilon$}
\put(0.0,4.55){$10^{-1}$}
\put(0.0,3.2){$10^{-2}$}
\put(0.0,1.85){$10^{-3}$}
\put(5.3,3.2){$\alpha_{\rm ex}-\alpha_N$}
\put(3.8,2.2){$\alpha^{\rm ex}_{\rm K}-\alpha^N_{\rm K}$}
\end{picture}
}
\caption[]{\label{fig1} Difference between the extrapolated pressure constant
$\alpha_{\rm ex}$ and the optimized
$N$-th order value $\alpha_N$ obtained from variational perturbation theory
for the method presented in this paper (solid line) and the first four
values of the
approximation scheme introduced in Ref.~\cite{kl277} (dashed line).
Dots represent the values to order $N$ in these approximations.}
\end{figure}

The approach is graphically shown in Fig.~\ref{fig1} where the optimized
values $\alpha_0,\ldots,\alpha_3$ all lie on a straight line (solid line).
For comparison, we have also extrapolated the first four
values $\alpha_{\rm K}^0,\ldots,\alpha_{\rm K}^3$ in the approach of
Ref.~\cite{kl277} yielding a value
$\alpha_{\rm ex\, K}\approx 0.0759786$, which is 4.9\%
smaller than (\ref{mem35}).
\section{Summary}
We have calculated the universal constant $\alpha$ occuring in the pressure
law (\ref{mem01}) of a membrane fluctuating between two walls. This has been
done by replacing the walls by a smooth potential with a parameter $m^2$.
This potential approaches the wall potential in the limit $m^2\to 0$.
The anharmonic part of the smooth potential was treated perturbatively.
The limit $m^2\to 0$ corresponds
to a strong-coupling limit of the power series, and was
calculated by
variational perturbation theory. Extrapolating the lowest four approximations
to infinity yileds a pressutre constant $\alpha$, which is in
very good agreement with Monte-Carlo values.
\begin{table}[p]
\caption[]{\label{tab1} Feynman dia\-grams with loops $L$,
multiplicities $s$,
and their dimensionless values $v$.
The last column shows the values $v_{\rm K}=v_{\rm PB}/4^L$ used in
Ref.~\cite{kl277}.}
\begin{tabular}{lcccc}
$L$ & Graph & $s$ & $v$ & $v_{\rm K}$  \\ \hline\hline
2 & \parbox{15mm}{\vspace{2mm}\centerline{
\begin{fmfgraph}(10,5)
\setval
\fmfforce{0.5w,0.5h}{v}
\fmfi{plain}{fullcircle scaled 1h shifted (1/4w,1/2h)}
\fmfi{plain}{fullcircle scaled 1h shifted (3/4w,1/2h)}
\fmfdot{v}
\end{fmfgraph}
}\vspace{2mm}} & $3$ & $\displaystyle{\frac{3}{64}}$ &
$\displaystyle{\frac{3}{64}}$\\ \hline
&\rule[-5pt]{0pt}{15pt} & & $a_1=a_1^{\rm K}$ &
$a_1^{\rm K}=1/64$\\ \hline\hline
3 & \parbox{15mm}{\vspace{2mm}\centerline{
\begin{fmfgraph}(10,7.5)
\setval
\fmfforce{0.5w,1/3h}{v}
\fmfi{plain}{fullcircle xscaled 0.3h yscaled 10/15h rotated 240
shifted (0.72w,0.15h)}
\fmfi{plain}{fullcircle xscaled 0.3h yscaled 10/15h rotated 120
shifted (0.28w,0.15h)}
\fmfi{plain}{fullcircle xscaled 0.3h yscaled 10/15h shifted (0.5w,0.67h)}
\fmfdot{v}
\end{fmfgraph}
}\vspace{1mm}} & $15$ & $\displaystyle{\frac{15}{512}}$ &
$\displaystyle{\frac{15}{512}}$\\
 & \parbox{20mm}{\vspace{1mm}\centerline{
\begin{fmfgraph}(15,5)
\setval
\fmfforce{1/3w,1/2h}{v1}
\fmfforce{2/3w,1/2h}{v2}
\fmf{plain,left=1}{v1,v2,v1}
\fmfi{plain}{fullcircle scaled 1/3w shifted (1/6w,0.5h)}
\fmfi{plain}{fullcircle scaled 1/3w shifted (5/6w,0.5h)}
\fmfdot{v1,v2}
\end{fmfgraph}
}\vspace{1mm}} & $72$ & $\displaystyle{\frac{9}{128}}$ &
$\displaystyle{\frac{9}{128}}$ \\
 & \parbox{20mm}{\vspace{1mm}\centerline{
\begin{fmfgraph}(7.5,7.5)
\setval
\fmfforce{0w,1/2h}{v1}
\fmfforce{1w,1/2h}{v2}
\fmf{plain,left=1}{v1,v2,v1}
\fmf{plain,left=0.5}{v1,v2,v1}
\fmfdot{v1,v2}
\end{fmfgraph}
}\vspace{2mm}} & $24$ & $0.828571\times \displaystyle{\frac{3}{256}}$ &
$\displaystyle{\frac{3}{256}}$\\ \hline
&\rule[-5pt]{0pt}{15pt} & & $a_2=1.114286\,a_2^{\rm K}$ &
$a_2^{\rm K}=1/1024$\\ \hline\hline
4 & \parbox{20mm}{\vspace{2mm}\centerline{
\begin{fmfgraph}(7.5,7.5)
\setval
\fmfforce{0.5w,0.5h}{v}
\fmfi{plain}{fullcircle xscaled 5/20h yscaled 10/15h rotated 225
shifted (0.75w,0.25h)}
\fmfi{plain}{fullcircle xscaled 5/20h yscaled 10/15h rotated 135
shifted (0.25w,0.25h)}
\fmfi{plain}{fullcircle xscaled 5/20h yscaled 10/15h rotated 45
shifted (0.25w,0.75h)}
\fmfi{plain}{fullcircle xscaled 5/20h yscaled 10/15h rotated -45
shifted (0.75w,0.75h)}
\fmfdot{v}
\end{fmfgraph}
}\vspace{1mm}} & $105$ & $\displaystyle{\frac{105}{4096}}$ &
$\displaystyle{\frac{105}{4096}}$ \\
 & \parbox{20mm}{\vspace{1mm}\centerline{
\begin{fmfgraph}(15,7.5)
\setval
\fmfforce{1/3w,1/2h}{v1}
\fmfforce{2/3w,1/2h}{v2}
\fmf{plain,left=1}{v1,v2,v1}
\fmfi{plain}{fullcircle scaled 1/3w shifted (5/6w,0.5h)}
\fmfi{plain}{fullcircle xscaled 5/20h yscaled 1/3w rotated 135
shifted (0.2w,0.25h)}
\fmfi{plain}{fullcircle xscaled 5/20h yscaled 1/3w rotated 45
shifted (0.2w,0.75h)}
\fmfdot{v1,v2}
\end{fmfgraph}
}\vspace{1mm}} & $540$ & $\displaystyle{\frac{135}{2048}}$ &
$\displaystyle{\frac{135}{2048}}$ \\
 & \parbox{20mm}{\vspace{1mm}\centerline{
\begin{fmfgraph}(12.5,7.5)
\setval
\fmfforce{10/25w,1/2h}{v1}
\fmfforce{1w,1/2h}{v2}
\fmf{plain,left=1}{v1,v2,v1}
\fmf{plain,left=0.5}{v1,v2,v1}
\fmfi{plain}{fullcircle scaled 10/25w shifted (5/25w,0.5h)}
\fmfdot{v1,v2}
\end{fmfgraph}
}\vspace{1mm}} & $360$ & $0.828571\times \displaystyle{\frac{45}{2048}}$ &
$\displaystyle{\frac{45}{2048}}$ \\
 & \parbox{20mm}{\vspace{1mm}\centerline{
\begin{fmfgraph}(20,7.5)
\setval
\fmfforce{1/4w,1/2h}{v1}
\fmfforce{2/4w,1/2h}{v2}
\fmfforce{3/4w,1/2h}{v3}
\fmf{plain,left=1}{v1,v2,v1}
\fmf{plain,left=1}{v2,v3,v2}
\fmfi{plain}{fullcircle scaled 1/4w shifted (5/40w,0.5h)}
\fmfi{plain}{fullcircle scaled 1/4w shifted (35/40w,0.5h)}
\fmfdot{v1,v2,v3}
\end{fmfgraph}
}\vspace{1mm}} & $2592$ & $\displaystyle{\frac{81}{512}}$ &
$\displaystyle{\frac{81}{512}}$\\
& \parbox{20mm}{\vspace{1mm}\centerline{
\begin{fmfgraph}(13.66,12.5)
\setval
\fmfforce{8.995/13.66w,3.75/12.5h}{v1}
\fmfforce{4.665/13.66w,3.75/12.5h}{v2}
\fmfforce{1/2w,7.5/12.5h}{v3}
\fmf{plain,left=0.6}{v1,v2,v3,v1}
\fmfi{plain}{fullcircle scaled 5/13.66w shifted (1/2w,10/12.5h)}
\fmfi{plain}{fullcircle scaled 5/13.66w shifted (11.15/13.66w,2.5/12.5h)}
\fmfi{plain}{fullcircle scaled 5/13.66w shifted (2.5/13.66w,2.5/12.5h)}
\fmfdot{v1,v2,v3}
\end{fmfgraph}
}\vspace{1mm}} & $1728$ & $\displaystyle{\frac{81}{512}}$ &
$\displaystyle{\frac{81}{512}}$\\
 & \parbox{20mm}{\vspace{1mm}\centerline{
\begin{fmfgraph}(12.5,7.5)
\setval
\fmfforce{10/25w,1/2h}{v1}
\fmfforce{8.75/12.5w,1h}{v2}
\fmfforce{8.75/12.5w,0h}{v3}
\fmf{plain,left=0.45}{v1,v2}
\fmf{plain,left=1}{v2,v3}
\fmf{plain,left=0.45}{v3,v1}
\fmf{plain,left=0.5}{v2,v3,v2}
\fmfi{plain}{fullcircle scaled 10/25w shifted (5/25w,0.5h)}
\fmfdot{v1,v2,v3}
\end{fmfgraph}
}\vspace{1mm}} & $3456$ & $0.828571\times \displaystyle{\frac{135}{1024}}$ &
$\displaystyle{\frac{135}{1024}}$ \\
& \parbox{20mm}{\vspace{1mm}\centerline{
\begin{fmfgraph}(7.5,7.5)
\setval
\fmfforce{0.502/7.5w,1.875/7.5h}{v1}
\fmfforce{6.998/7.5w,1.875/7.5h}{v2}
\fmfforce{1/2w,1h}{v3}
\fmf{plain,right=0.55}{v1,v2,v3,v1}
\fmf{plain}{v1,v2}
\fmf{plain}{v2,v3}
\fmf{plain}{v3,v1}
\fmfdot{v1,v2,v3}
\end{fmfgraph}
}\vspace{2mm}} & $1728$ & $0.713194\times \displaystyle{\frac{81}{2048}}$ &
$\displaystyle{\frac{81}{2048}}$\\ \hline
&\rule[0pt]{0pt}{10pt} & & $a_3=2.763097\cdot 10^{-5}$ & $a_3^{\rm K}=0$
\end{tabular}
\end{table}
\end{fmffile}
\end{document}